\documentclass{ws-procs9x6}
\usepackage{psfrag}

\newcommand{\xrightarrow}[2][]{
  \mathrel{\mathop{%
    \setbox0\hbox{$\rightarrow$}%
    \setbox2\vbox{
      \hbox{$\scriptstyle#2$}%
      \copy0
    }%
      \raisebox{0.6ex}{\hbox to0pt{\hbox to\wd2{\hfil$\scriptstyle #2$\hfil}}}%
      \raisebox{-0.4ex}{\hbox to \wd2{\hfil$\rightarrow$\hfil}}%
}}}
\begin{document}

\title{Indirect Detection of CMSSM Neutralino Dark Matter with Neutrino
  Telescopes} 
\author{J. Orloff\footnote{Presented by \uppercase{J}.
    \uppercase{O}rloff} \ and E. Nezri}

\address{Laboratoire de Physique Corpusculaire, \\
IN2P3-CNRS, Universit\'e Blaise Pascal, F-63177 Aubi\`ere Cedex\\
E-mail: orloff@in2p3.fr, nezri@in2p3.fr}

\author{V. Bertin}

\address{Centre de Physique des Particules de Marseille, \\
IN2P3-CNRS, Universit\'e de la M\'editerran\'ee, F-13288 Marseille Cedex 09\\
E-mail: bertin@in2p3.fr}  


\maketitle

\abstracts{We review the prospects of detecting supersymmetric dark matter
  in the framework of the Constrained Minimal Supersymmetric Standard
  Model, and compare indirect with direct detection capabilities.}

\begin{figure}[t]
\psfrag{t+t-}[l][l]{$\tau^+\tau^-$}
\psfrag{W+W-}[l][l]{$W^+W^-$}
\psfrag{1}[l][l]{1}
\psfrag{1e-1}[l][l]{$10^{-1}$}
\psfrag{1e-2}[l][l]{$10^{-2}$}
\psfrag{1e-3}[l][l]{$10^{-3}$}
\psfrag{1e-4}[l][l]{$10^{-4}$}
\psfrag{neutrino energy}[l][l]{Neutrino energy}
\psfrag{M}[l][l]{$M$}
\psfrag{c}[l][l]{$\chi$}
\psfrag{1/4}[l][l]{1/4}
\psfrag{1/2}[l][l]{1/2}
\psfrag{3/4}[l][l]{3/4}
\psfrag{Differential neutrino spectra}[l][l]{Differential neutrino spectra}
\psfrag{f\(n\)}[l][l]{$\frac{d\phi(E)}{dE}$}
\centerline{\includegraphics[width=0.85\textwidth]{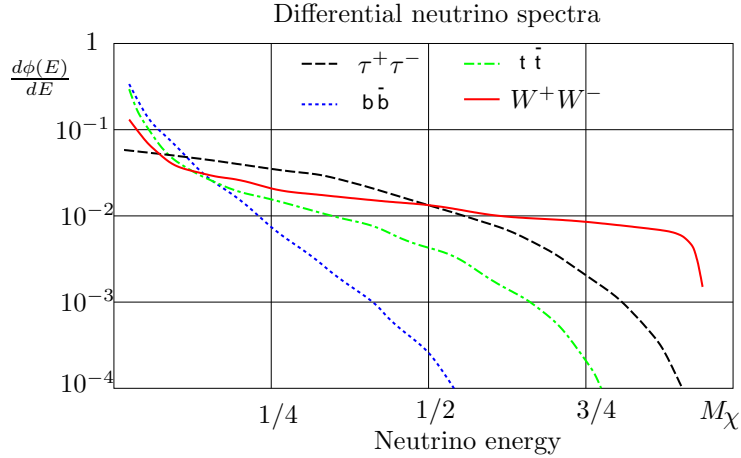}}
\caption{Energy dependence of secondary neutrino fluxes (in arbitrary
  units) from dominant
  neutralino annihilation channels, as functions of the neutrino energy in
  units of the neutralino mass.}
\label{nufluxdiff}
\end{figure}

Recently, both theoretical considerations and and a wealth of experimental
data in cosmology have converged towards a $\Lambda CDM$ flat and black
universe, with the following amounts of dark energy and cold dark matter:
$\Omega_{\Lambda}\sim 0.7$, $\Omega_{CDM}\sim 0.3$. This last fraction
could be incarnated by a bath of Weakly Interacting Massive Particles
(WIMPs), whose annihilation stopped when the universe expansion separated
them enough from each other, leaving a non relativistic relic density.  In
the MSSM framework, assuming R-parity conservation, the Lightest
Supersymmetric Particle (LSP) is stable and is the lightest neutralino
($\doteq$ {\it the} neutralino(s) $\chi$) in most regions of the parameter
space. If present in galactic halos, relic neutralinos must accumulate in
astrophysical bodies (of mass $M_b$) like the Earth or most importantly the
Sun\cite{Jungman:1996df}, which then play the role of cosmic storage rings
for neutralinos. The capture rate $C$ depends on the neutralino-quark
elastic cross section: $\sigma_{\chi-q}$.  Neutralinos being Majorana
particles, their vectorial interaction vanishes and the allowed
interactions are scalar (via $ \chi q \xrightarrow{H,h} \chi q$ in $t$
channel and $ \chi q \xrightarrow{\tilde{q}} \chi q$ in $s$ channel) and
axial (via $\chi q \xrightarrow{Z} \chi q$ in $t$ channel and $ \chi q
\xrightarrow{\tilde{q}} \chi q$ in $s$ channel). Depending on the spin
content of the nuclei $N$ present in the body, scalar and/or axial
interactions are involved.  Roughly, $C\sim \frac{\rho_{\chi}}{v_{\chi}}
\sum_N M_b f_N \frac{\sigma_N}{m_{\chi}m_N} <v^2_{esc}>_N
F(v_{\chi},v_{esc},m_{\chi},m_N)$, where $\rho_{\chi}, v_{\chi}$ are the
local neutralino density and velocity, $f_N$ is the density of nucleus $N$
in the body, $\sigma_N$ the nucleus-neutralino elastic cross section,
$v_{esc}$ the escape velocity and $F$ a suppression factor depending on
masses and velocity mismatching.  Considering that the population of
captured neutralinos has a velocity lower than the escape velocity, and
therefore neglecting evaporation, the total number $N_{\chi}$ of
neutralinos in a massive astrophysical object depends on the balance
between capture and annihilation rates: $\dot{N_{\chi}}=C-C_AN_{\chi}^2$,
where $C_A$ is the total annihilation cross section $\sigma^A_{\chi-\chi}$
times the relative velocity divided by the volume. The annihilation rate at
a given time $t$ is then:
\begin{equation}
\Gamma_A=\frac{1}{2}C_AN_{\chi}^2=\frac{C}{2}\tanh^2(t\sqrt{CC_A})
\label{eq:annihirate}
\end{equation}
with $\Gamma_A\approx \frac{C}{2}=cste$ when the neutralino population has
reached equilibrium, and $\Gamma_A\approx \frac{1}{2}C^2C_At^2$ in the
initial collection period (relevant in the Earth). So, when accretion is
efficient, the annihilation rate does not depend on annihilation processes
but follows the capture rate $C$ and thus the neutralino-quark elastic
cross section. The neutrino differential flux resulting from $\chi\chi$
annihilation is given by:
\begin{equation}
\frac{d\Phi}{dE}= \frac{\Gamma_A}{4 \pi R^2} \sum_F B_F 
                  \left(\frac{dN}{dE}\right)_{F}
\label{eq:nuflux}
\end{equation}
where $R$ is the distance between the source and the detector, $B_F$ is the
branching ratio of annihilation channel $F$ and $(dN/dE)_F$ its differential
neutrino spectrum.  As the direct neutrino production $\chi\chi\rightarrow
\nu\bar{\nu}$ exactly vanishes in the massless neutrino limit, neutrino
fluxes mainly come from decays of primary annihilation products, with a
mean energy $E_{\nu}\sim\frac{m_{\chi}}{2}$ to $\frac{m_{\chi}}{3}$ (see
figure~\ref{nufluxdiff}). The most energetic ``hard'' spectra come from
neutralino annihilations into $WW$ or $ZZ$, and the less energetic ``soft''
ones come from $b\bar{b}$. Neutrino telescopes use the Earth as a target
for converting the muon component of these neutrino fluxes into measurable
muons (see S. Cartwright, these proceedings). As both the
$\nu_\mu$ charged-current cross section on Earth nuclei and the produced muon
range are proportional to $E_{\nu}$, high energy neutrinos are easier to
detect.

The branching fractions $B_F$ are thus relevant neutralino properties that
depend on the particular SUSY model considered.  We\cite{Bertin:2002ky}
have studied these in the Constrained Minimal Supersymmetric Standard Model
(CMSSM, {\it a.k.a.}  mSugra), whose attractiveness comes from a tractable
number of free parameters: $m_0$ (common scalar mass), $m_{1/2}$ (common
gaugino mass), $A_0$ (common trilinear term) and $sign(\mu)$
(supersymmetric scalar mass term), all fixed at a high energy scale
$E_{GUT}\sim2.10^{16}$ GeV, as well as $\tan\beta$, fixed at the EW scale.
The coexistence of these widely different scales introduces theoretical
uncertainties on the exact definition of the model (especially at large
$\tan\beta$), but the advent of more and more reliable Renormalization
Group Equations codes (like {\tt Suspect2.005}\cite{Djouadi:2002ze} used in
this work) tends to reduce these. As a bonus, coping with RGE's from the
start guarantees the expandability of the model to high energies which is
the main motivation for introducing SUSY and neutralinos in the first
place.

\begin{figure}[t]
  \centering
\includegraphics[width=\textwidth]{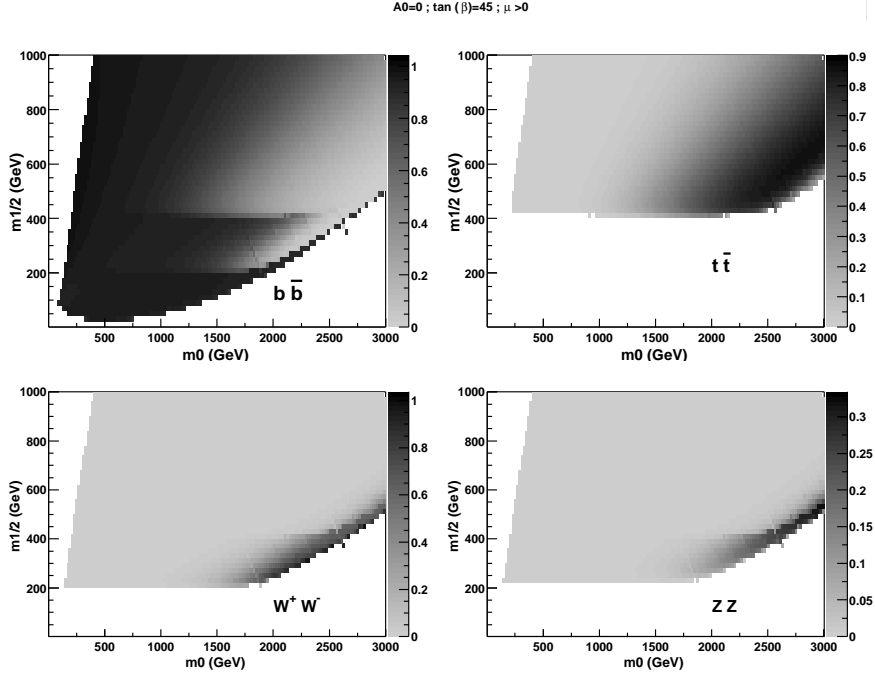}  
  \caption{The dominant neutralino annihilation branching ratios for a
    typical large $\tan\beta=45$ case, as functions of $m_0$ (common scalar
    mass) and $m_{1/2}$ (common gaugino mass).}
  \label{bratio}
\end{figure}

As seen on figure~\ref{bratio}, the hard spectra from $W^+W^-$ and $t\bar
t$, are found at large $m_0$ for fixed $m_{1/2}$ larger than the
corresponding threshold. In this ``focus point'' region\cite{Feng:2000gh},
the neutralino has a sizeable higgsino component $h_{frac}(\chi^0)$ which
allows its annihilation into gauge bosons via $t$-channel gaugino exchange,
with a cross-section $\sigma_A\propto h_{frac}^2(\chi^0)
h_{frac}^2(\chi^+)$ and an interesting relic density may survive. Although
this region seems a small fine-tuned corner of the $(m_0,m_{1/2})$ plane,
relaxing universality may help in this
respect\cite{Bertin:2002sq,Ellis:2002wv} Otherwise, the
neutralino is an almost pure bino mainly annihilating into $b\bar b$
through $s$-channel $A$ exchange or $t$-channel sfermion exchange; a low
enough relic density can only be found at small $m_0$ for fixed but not too
large $m_{1/2}$.

To study the detectability of mSugra dark matter, we have used the {\tt
  DarkSusy}\cite{Darksusy} code and computed 1) the relic density, 2) the
solar muon flux seen by neutrino telescopes and 3) the scalar elastic cross
section $\sigma^{scal}_{\chi-p}$ relevant to Germanium or Xenon direct
detection, for a wide range of such mSugra models: $m_{1/2}\in(50,\, 1000)$
GeV, $m_0\in (0,\, 3000)$ GeV, $tan\beta=10,20,35,45,50$, $\mu>0$,
$A_0=-800,-400,0,400,800$ GeV (for $tan\beta=20,35$ only). Among these, we
kept only those satisfying the following accelerator constraints:
$BR(b\rightarrow s\gamma)\in(2.2,\, 5.2) \times 10^{-4}$,
$a_\mu^{susy}\in(-6,\, 58) \times 10^{-10}$, $m_{\chi_1^+}>104$ GeV,
$m_{h}> 113$ GeV.

\begin{figure}[t]
  \centering
\includegraphics[width=0.495\textwidth]{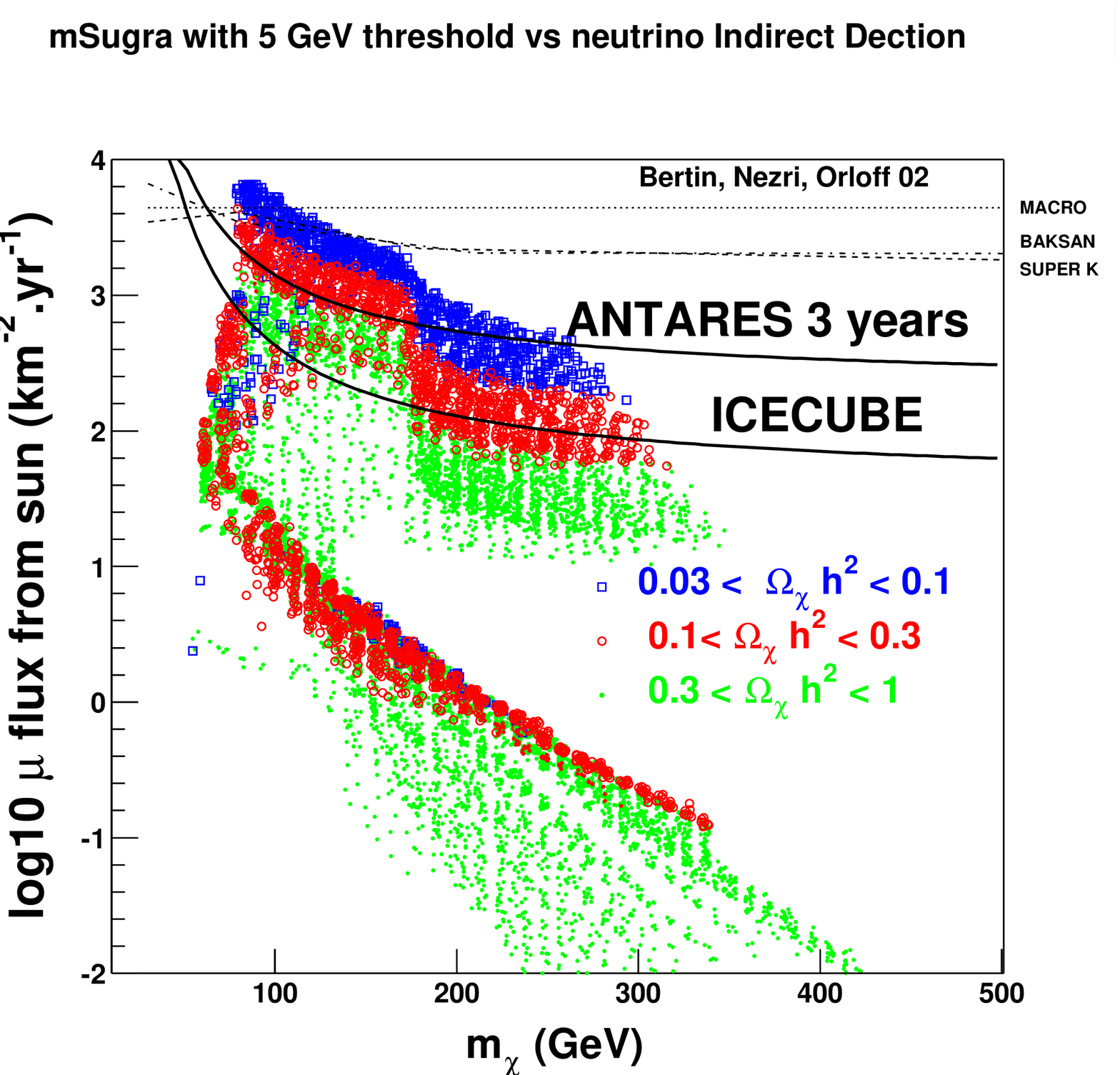}
\includegraphics[width=0.495\textwidth]{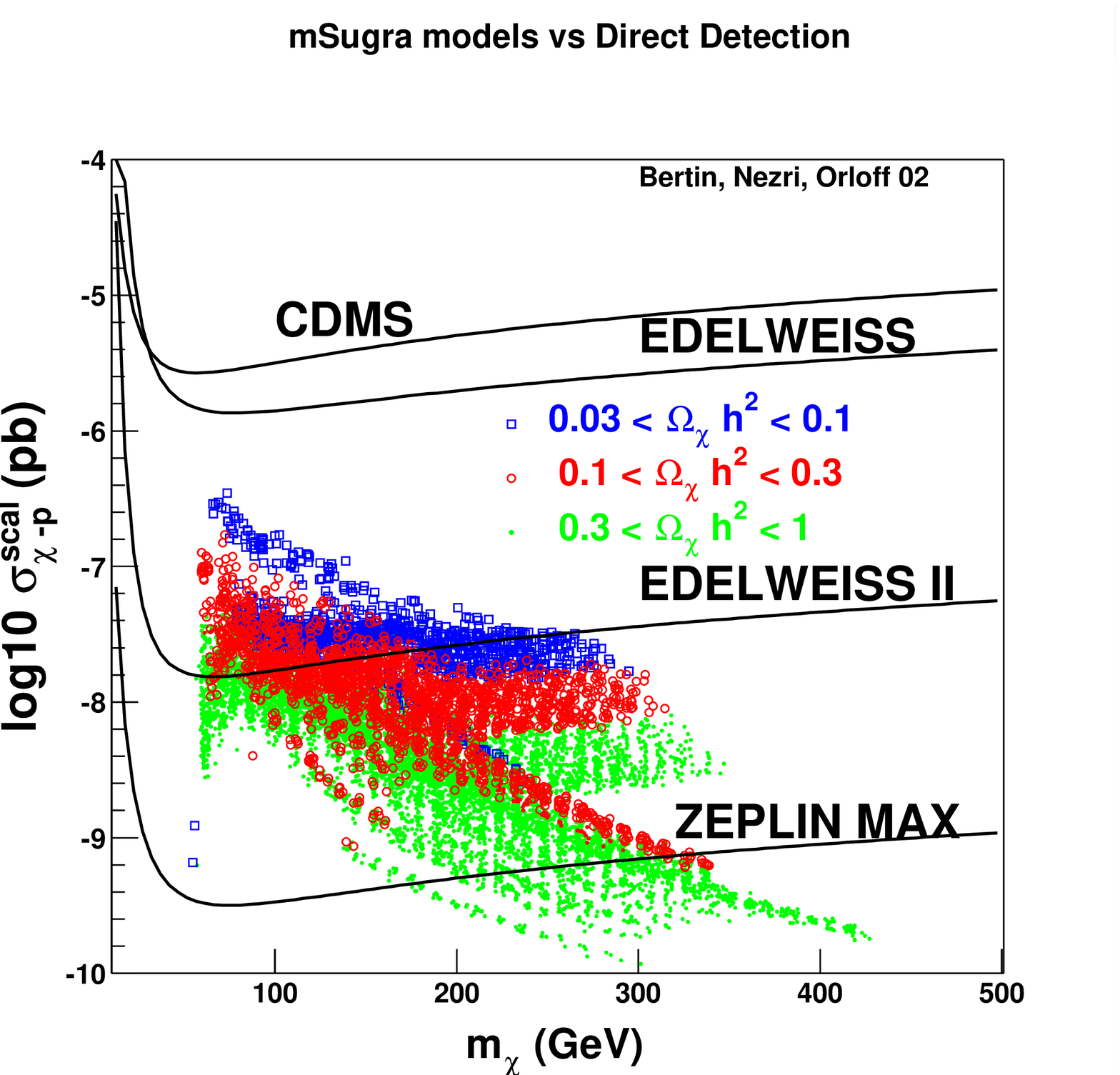}
  \caption{Indirect (left) and direct (right) detection potential of
    present and future experiments for mSugra models satisfying
    present experimental limits and offering interesting relic densities.}
  \label{in-vs-direct}
\end{figure}

In left figure~\ref{in-vs-direct}, these models are sorted according to
their neutralino mass and the muon flux $\Phi_{\mu}$ above 5GeV 
originating from the Sun. This is compared with past and future
experimental sensitivities assuming the hardest neutrino spectrum of
figure~\ref{nufluxdiff} normalized to $\Phi_{\mu}$: the lower
threshold of Baksan or SuperK thus show as a better sensitivity at low
$m_\chi$. Applying the shown cuts on the relic density\footnote{The
  dimensionless Hubble parameter squared $h^2$ is about 0.5} separates the
models in the 2 rough classes indicated above: the lower half corresponds
to binos, while the upper half is populated by models in the ``focus
point'' region and neutralinos with a sizeable $h_{frac}(\chi)$. In this
region, one clearly notices the $W^+W^-$ and $t\bar t$ thresholds at
$m_\chi=89$ and 175 GeV respectively. Between these, the neutrino spectrum
is indeed hard, and we see that Antares has the potential of detecting the
models with the expected relic density $\Omega=0.3$. For fixed $m_\chi$,
one also notices the correlation $\Phi_\mu\propto(\Omega h^2)^{-1}$, which
can be understood as both the annihilation amplitude (determining the relic
density) and the spin dependent collision amplitude (determining the
capture in the Sun and thus the muon flux) are $\propto h_{frac}^2(\chi)$.
When the mSugra neutralino is a bino, its spin dependent capture in the Sun
is much reduced and the muon flux is far below present or future
detection abilities. Similarly, neutralinos captured and annihilating in
the Earth give far too low fluxes for mSugra models.

Turning to direct detection, the right figure~\ref{in-vs-direct} shows that for
small masses, both the bino and focus region neutralinos are within reach
of the next generation of direct detection experiments like Edeweiss II.
The smaller vertical spread can be traced to the fact that the spin
independent (or scalar) collision amplitude is proportional to only one power of
$h_{frac}(\chi)$, which results in a much weaker increase in the focus
region. Also notice that an ultimate direct detection tool like Zeplin in a
maximal version seems to cover all interesting relic densities. However
coannihilations with staus (not included here) can allow for larger masses
which would still be out of reach.

\begin{figure}[t]
  \centering
  \includegraphics[width=0.49\textwidth]{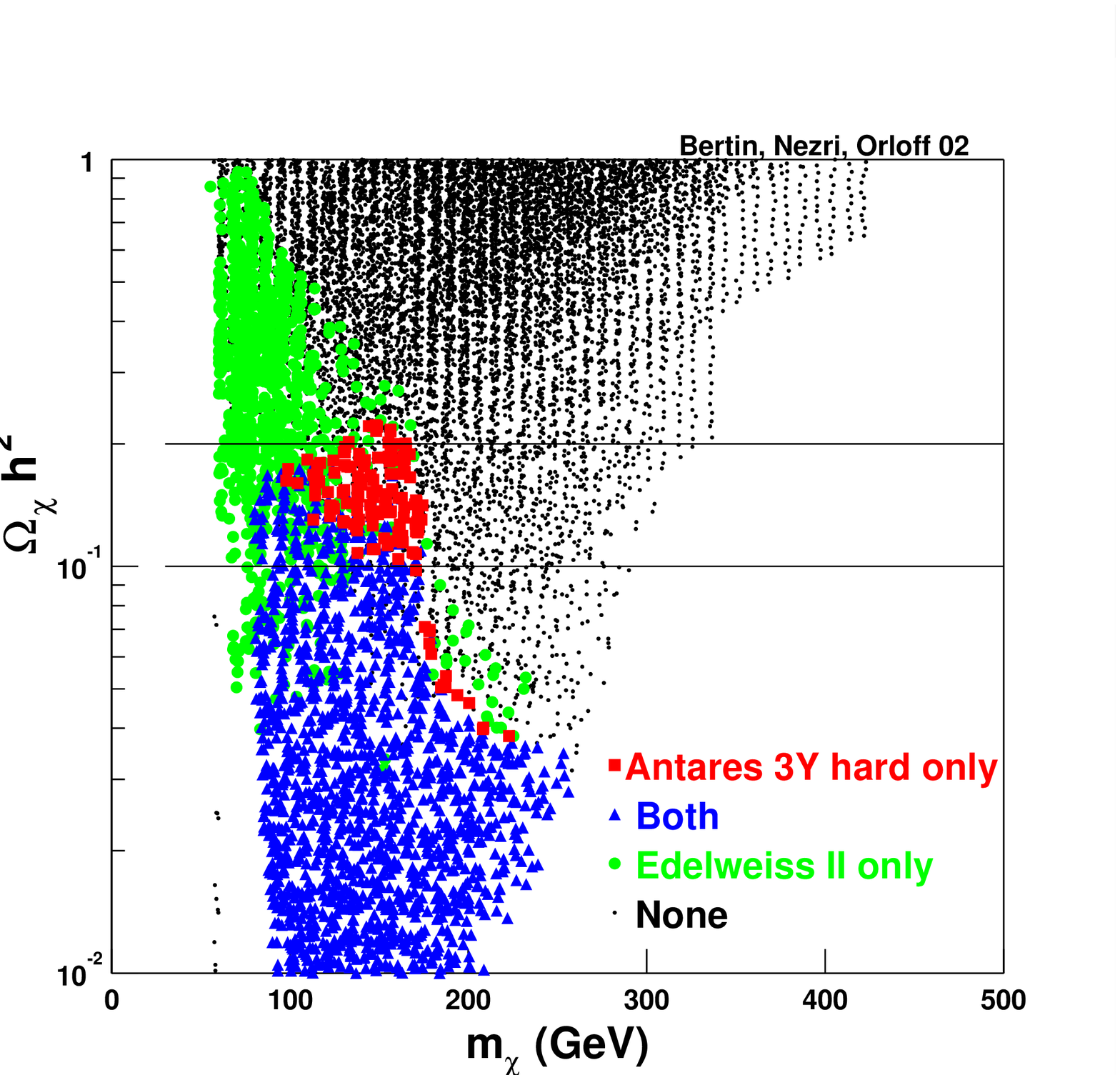}
  \includegraphics[width=0.49\textwidth]{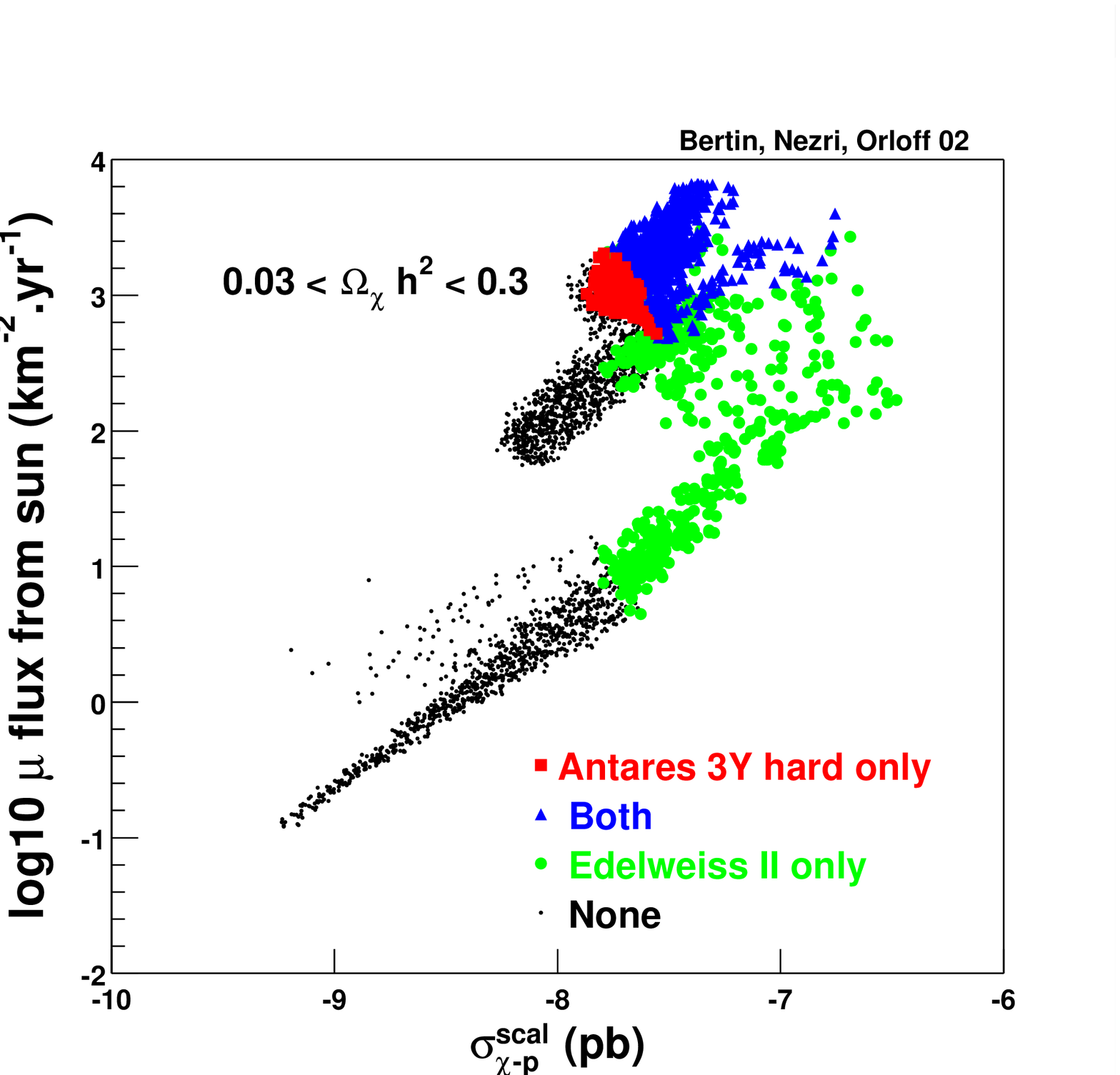}
  \caption{Other projections of the same mSugra models as in 
    fig~\ref{in-vs-direct}: left in the $(m_\chi,\Omega h^2)$ plane; right in the
    $(\sigma^{scalar}_{\chi-p},\Phi_{\mu})$ plane for $\Omega h^2 \in
    (0.03,0.3)$. Each model is labelled according to its detectability: by
    Antares only, by Edelweiss II only, by both or none.  }
  \label{flux-vs-coll}
\end{figure}

Another way to compare the merits of (in)direct detectors of
mSugra dark matter is shown on figure~\ref{flux-vs-coll}. On the left, all
mSugra models of the set defined above are placed in the $(m_\chi,\Omega
h^2)$ plane and sorted by their detectability. On the right, the
models with a relic density $\Omega h^2\in[0.03,0.3]$ are placed in the
$(\sigma^{scalar}_{\chi-p},\Phi_{\mu})$ plane and sorted the
same. Notice again the split in 2 groups, the upper half one again being
the mixed neutralinos of the focus region. A complementarity between direct
and indirect detection emerges from this splitting.


\end{document}